\documentclass[epj]{svjour}

\usepackage{epsf}

\begin{document}

\newcommand{\ie}{\mbox{i.\ e.\ }}
\newcommand{\eg}{\mbox{e.\ g.\ }}
\newcommand{\nb}{\mbox{N.\ B.\ }}
\newcommand{\cf}{\mbox{c.\ f.\ }}
\newcommand{\etal}{\mbox{\it et.\ al.\ }}
\newcommand{\be}{\begin{eqnarray}}
\newcommand{\en}{\end{eqnarray}}
\newcommand{\no}{\nonumber}
\newcommand{\hc}{\mbox{\it h.\ c.\ }}
\newcommand{\mc}{\mathcal}
\newcommand{\half}{\frac{1}{2}}

\title{Mixed valence on a pyrochlore lattice --- $LiV_2O_4$
as a geometrically frustrated magnet.
}

\author{Nic Shannon} 
               
\institute{Max--Planck--Institut f{\"u}r Physik komplexer Systeme,
N{\"o}thnitzer Str. 38, 01187 Dresden, Germany. }
\date{Received: date / Revised version: date}

\abstract{
Above $40K$, the magnetic susceptibility of the heavy 
Fermion spinel $LiV_2O_4$ has many features in common
with those of geometrically frustrated magnetic 
insulators, while its room temperature resistivity 
comfortably exceeds the Mott--Regel limit. 
This suggests that local magnetic moments,
and the underlying geometry of the pyrochlore lattice,
play an important role in determining its magnetic
properties.
We extend a recently introduced tetragonal mean field
theory for insulating pyrochlore antiferromagnets
to the case where individual tetrahedra contain
spins of different lengths, and use this as a starting point 
to discuss three different scenarios for magnetic
and electronic transitions in $LiV_2O_4$.
}

\maketitle

\section{Introduction}
\label{introduction}

Geometrically frustrated magnetic insulators continue to 
fascinate experimental and theoretical physicist alike.  
These systems are intriguing 
because the physics of a wide range of 
materials, with an equally wide range of physical properties, 
is underpinned by alluringly simple considerations of 
symmetry and entropy.
Perversly, the properties of frustrated systems which are 
structurally far more complicated than ``textbook'' magnetic 
insulators can therefore 
sometimes be understood on the basis of very simple arguments.

Recently, the geometrically frustrated ``metal'' LiV$_2$O$_4$
has also attracted a great deal of attention as the first 
example of d--electron heavy Fermion system.
In this article we apply simple arguments borrowed
from the study of frustrated magnetic insulators
to the magnetic susceptibility of LiV$_2$O$_4$
over the temperature range 30--1000K.
We argue that our simple model provides 
a good starting point for understanding the role 
of local geometric effects in the physics of LiV$_2$O$_4$,
and use it to explore the strength and weaknesses
of three different scenarios for the magnetic ``transitions''
seen in this material.

Our analysis is based on the extension of a recently 
introduced tetragonal mean field theory to a system with
a mixture of different magnetic moments.  We neglect
the partially itinerant nature of d--electrons in LiV$_2$O$_4$.
This approximation limits the range of temperatures
over which the theory is valid, but can 
be justified on the basis of simple physical arguments.
This article is therefore divided into two parts.
In sections 2 to 4 we review and extend the mean field theory
for a magnetic insulator on a pyrochlore lattice.
In section 5 we apply the generalized theory to LiV$_2$O$_4$.
and discus the remaining puzzles presented by the magnetic 
susceptibility of this most unusual material.

\section{Model}
\label{model}

\subsection{The Heisenberg model on a pyrochlore lattice}

The usual starting point for understanding the physics of
magnetic insulators is the Heisenberg Hamiltonian 
\be
\label{eqn:heisenberg}
{\mc H} &=& \sum_{ij} J_{ij} \vec{S}_i . \vec{S}_j
\en
where $\vec{S}_i$ is the operator for the spin of 
electrons on site $i$, and the matrix element $J_{ij}$ 
describes the (super--)exchange interaction between electrons on 
sites $i$ and $j$, and may be positive (antiferromagnetic)
or negative (ferromagnetic).   In many cases interaction can be 
restricted to nearest neighbour terms $J_{\langle ij\rangle}$.

We consider the antiferromagnetic (AF) Heisenberg model 
with all $J_{\langle ij \rangle} > 0$,
on the geometrically frustrated pyrochlore lattice.
This is a (sub--)structure common to 
many different magnetic insulators 
and also to a number of metallic systems, including  
the magnetically active V sites of the spinel LiV$_2$O$_4$.
Antiferromagnetic nearest neighbour exchange 
interactions favour anti--parallel spin alignments.   For a 
bipartite lattice this presents no problem, and the classical 
groundstate of equation~(\ref {eqn:heisenberg}) is the N\'eel State where
each sublattice has its maximal spin, and the two sublattices are
aligned anti--parallel to one another, so that the system has no net
spin, and each bond between spins has its lowest possible energy.
However the pyrochlore lattice falls into a more general 
class of lattices which exhibit an effect known as 
geometric frustration --- it is impossible to construct a classical
spin configuration in which all neighbouring spins are aligned
anti--parallel to one another.   
Where this is the case, many different states can become 
degenerate, and geometrically frustrated magnets therefore tend
to have a high (classical) ground state degeneracy.   The many
different degenerate classical groundstates are connected by 
operations reflecting the underlying symmetry of the lattice, 
and at a classical level this leads to the existence of branches of 
zero energy excitations in addition to the expected goldstone
modes of the system. 

Quantum and/or thermal fluctuations may enable a frustrated
system to chose its true groundstate by lifting the degeneracy 
between different classical spin configurations (equivalently,
generating a mass for all unphysical zero energy excitations).
This effect is known as ``order from disorder'', following 
a classic paper by Villain \cite{villain}, but calculations of order
from disorder effects in quantum mechanical spin systems 
based on large $S$ or large $N$ expansions must take proper 
account of zero energy modes, and are usually very involved
(see \eg \cite{andrey}, or for a recent example involving
itinerant electrons \cite{shannon}).  
It is desirable therefore to find a more economical way of 
calculating the properties of such systems.  One way to do 
so is to start in a basis of states which already reflects
the local symmetries of the lattice.   

\begin{figure}[tb]
\begin{center}
\leavevmode
\epsfxsize \columnwidth
\epsffile{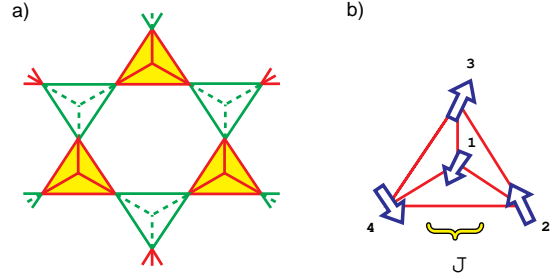}
\caption{a) Section of pyrochlore lattice showing two sublattice 
structure in terms of opposing tetrahedra.   Solid tetrahedra
point out of the plane, unfilled tetrahedra point into the plane.
b) Spins are found at the corners of tetrahedra, and shared
between $A$ and $B$ sublattices.}
\label{fig1}
\end{center}
\end{figure}

While the pyrochlore lattice has an overall cubic symmetry,  
in terms of the bonds between magnetically active sites,
it may be thought of as two inter--penetrating sublattices of
tetrahedra, with a spin at the corner of each tetrahedron.
Each spin is shared between a tetrahedron in the $A$ and a
tetrahedron in the $B$ sublattice.
The bases of both $A$ and $B$ sublattice tetrahedra lie in planes, 
and the bonds within these planes form a Kagom\'e lattice.
If we consider a given plane, the tetrahedra of one sublattice
will point into that plane, and those of the other sublattice out 
of it.  Neighbouring planes are joined by pairs of opposing tetrahedra.   
An illustration of this structure is shown in {\mbox figure~\ref{fig1}}.
Since individual spins are shared between the $A$ and $B$ sublattices, 
we may completely specify the state of the system by specifying the 
spin configurations of the tetrahedra on one sublattice.  We will
call this the $A$ sublattice.   Furthermore, if we neglect all bonds 
belonging to the $B$ sublattice, the $A$ sublattice reduces to a set of 
independent tetrahedra.   These independent tetrahedra will form the 
basis for our mean field theory.

\subsection{An individual tetrahedral subunit}

We now consider an individual tetrahedron on the $A$ sublattice,
described by 
\be
\label{eqn:tetrahedron}
{\mc H}_{TET} &=& {\mc H}_{EX} + {\mc H}_{h}
\en
where ${\mc H}_{EX}$ the Heisenberg Hamiltonian
\be
\label{eqn:exchange}
{\mc H}_{EX} &=& \sum_{\langle ij\rangle_{Tet} } 
   J_{ij}\vec{S}_i . \vec{S}_j
\en
and in order to calculate the susceptibility we 
introduce an external magnetic field along the z--axis
\be
{\mc H}_{h} &=& h \sum_i S_i^z
\en
Here the indices $i$,$j$ denote sites at different corners of the 
tetrahedron, and the sum is restricted so as to count each bond between 
spins only once.

\begin{figure}[tb]
\begin{center}
\leavevmode
\epsfysize = 120.0mm
\epsffile{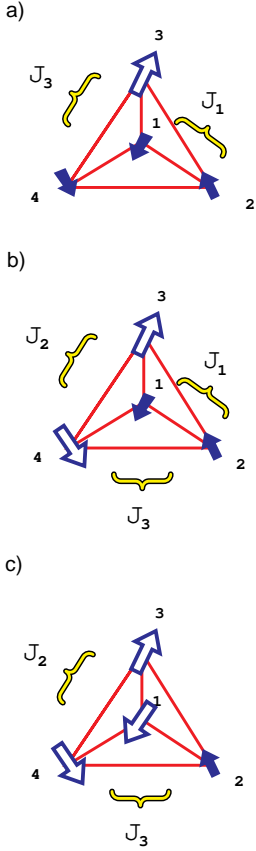}
\caption{
Mixed spin tetrahedra with 1, 2 and 3 spin 1, 
showing the different Heisenberg couplings 
J$_1$,J$_2$ and J$_3$.}
\label{fig4}
\end{center}
\end{figure}

This subunit is a system of four interacting local moments, and we 
consider the case in which the magnetic ions at each corner of the 
tetrahedron may take on one of two possible values of total spin,
these being either ``large'' (specifically, $S=1$ in what follows), 
or ``small'' (below, $s=1/2$).
The exchange integral $J_{ij}$ will in general vary
with the size of the spins at sites $i$ and $j$.
We use the notation $J_1$ to refer to the exchange interaction
between two small spins,  $J_2$ to the exchange interaction
between two large spins, and $J_3$ to the interaction between
two spins of different size, as illustrated in figure~(\ref{fig4}). 
We consider only the case of antiferromagnetic (AF) interaction 
\ie all $J >0$.

Since any given tetrahedron may have 0,1,2,3 or 4 large spin moments
(the remainder being of the small spin) we must consider five different
cases.  We will not consider the 
additional charge degeneracy associated with the different ways
of distributing spins throughout the tetrahedron as, for an
insulator, this has no dynamics.

\subsubsection{Excitation spectrum and partition function}

Since total spin is conserved 
for any isolated tetrahedron, it must 
be possible to diagonalize the Hamiltonian~(\ref{eqn:exchange})
in the basis of eigenstates of total spin
\be 
\vec{\Omega} &=& \vec{S}_1 + \vec{S}_2 + \vec{S}_3 + \vec{S}_4
\en
and its z--component $\Omega^z$.  If we further introduce the 
total spin of the ``small'' and ``large'' spin subsystems
\be 
\vec{\sigma} 
   &=& \sum_{\{i\}_{\mc Tet}} \vec{S}_i
       \delta_{S_i \frac{1}{2}} \no\\
\vec{\Sigma} 
   &=& \sum_{\{i\}_{\mc Tet}} \vec{S}_i
       \delta_{S_i 1}
\en
the Hamiltonian~(\ref{eqn:exchange}) can be written 
\be 
\label{eqn:spectrumH}
{\mc H}_{Tet} 
   &=& \frac{1}{2} \left[
         J_{\Omega}\vec{\Omega}^2
         + J_{\sigma}\vec{\sigma}^2
         + J_{\Sigma}\vec{\Sigma}^2
       \right] + const.
\en
where $J_{\Omega} = J_3$, $J_{\sigma} = J_1 - J_3$ and 
$J_{\Sigma} = J_2 - J_3$.
The coupling to external magnetic field is now simply
$
{\mc H}_{h} = h \Omega^z
$
and the excitation spectrum of the model in the absence of 
a magnetic field can be read directly
from the Hamiltonian~(\ref{eqn:exchange})
\be
\label{eqn:spectrum}
E(\Omega,\sigma,\Sigma)
   &=&   \frac{1}{2} \left[
         J_{\Omega} \Omega(\Omega + 1)
         + J_{\sigma}\sigma(\sigma + 1)
      \right.  \no\\ && \quad \left.   
         + J_{\Sigma}\Sigma(\Sigma + 1)
       \right]
\en
The ground state of the tetrahedron will be a spin--singlet for 
all $J_{\Omega} > 0$, but may be degenerate for mixed spin systems.
Since we anticipate $J_2 > J_3 > J_1$, in general $J_{\sigma} < 0$
and the smaller spins tend to be aligned in order to collectively 
screen the larger ones.

In order to calculate the partition function of the 
tetrahedron, we also need to know the degeneracy $g(\Omega,\sigma,\Sigma)$
of each state.
We will not discuss the (tedious) details of the evaluation of 
these degeneracy factors, but note that they can be found using a simple 
generalization of the method introduced for systems with a single type 
of spin by van Vleck (see Appendix \ref{appendix1}).  
Actual degeneracies for the 
states of tetrahedra with no, one, two, three and four spin
$S=1$ spins are listed in Appendix \ref{appendix2}.

Given knowledge of $E(\Omega,\sigma,\Sigma)$ and 
$g(\Omega,\sigma,\Sigma)$, the 
partition function of the tetrahedral subunit in
the presence of a magnetic field $h$ at temperature $T$
can be expressed as
\be
Z &=& \sum_{\Omega\sigma\Sigma} g(\Omega,\sigma,\Sigma)
   \exp\left(-\frac{E(\Omega,\sigma,\Sigma)}{T}\right) 
      \no\\ && \qquad   
   \times F_{\Omega}\left(\frac{h\Omega}{T}\right)
\en 
where $F_{\Omega}(x)$ is the function
\be
F_{\Omega}(x) 
   &=& \frac{
   \sinh \left(\frac{(2\Omega+1)x}{2\Omega}\right)
   }{
   \sinh \left(\frac{x}{2\Omega}\right)
   }
\en      

\subsubsection{Entropy and Specific heat}

\begin{figure}[tb]
\begin{center}
\leavevmode
\epsfxsize \columnwidth
\epsffile{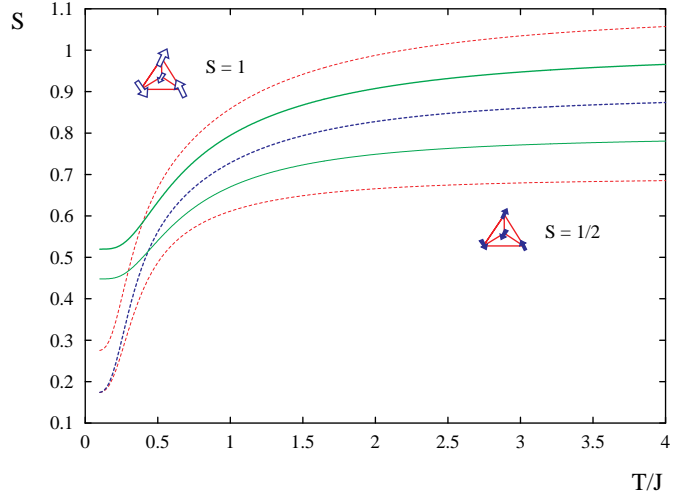}
\caption{
Entropy of spin $1/2$, spin $1$ and mixed
spin tetrahedra as a function of temperature
in units where $k_B=1$.
From top to bottom at RHS plot, 
tetrahedra with --- 
4 spin 1 (dotted line),
3 spin 1 and 1 spin 1/2 (solid line), 
2 spin 1 and 2 spin 1/2 (dotted line),
1 spin 1 and 3 spin 1/2 (solid line),
4 spin 1/2 (dotted line).
All couplings set equal to J.}
\label{fig2}
\end{center}
\end{figure}

\begin{figure}[tb]
\begin{center}
\leavevmode
\epsfxsize \columnwidth
\epsffile{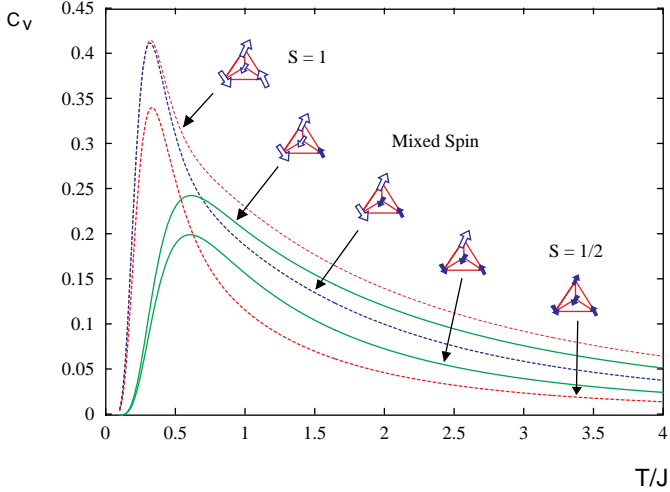}
\caption{Heat capacity of spin $1/2$, spin $1$ and mixed
spin tetrahedra as a function of temperature in units
where $k_B=1$.
From top to bottom at RHS plot, 
tetrahedra with --- 
4 spin 1 (dotted line),
3 spin 1 and 1 spin 1/2 (solid line), 
2 spin 1 and 2 spin 1/2 (dotted line),
1 spin 1 and 3 spin 1/2 (solid line),
4 spin 1/2 (dotted line).
All couplings set equal to J.}
\label{fig3}
\end{center}
\end{figure}

The entropy of an individual tetrahedral subunit is 
given by 
\be
S &=& \ln Z + \frac{\langle E \rangle}{T}
\en
where the average energy of the system $\langle E \rangle$ 
is 
\be
\langle E \rangle &=& \frac{1}{Z} \sum_n E_n e^{-\frac{E_n}{T}} 
\en
The sum over states $\{n\}$ involved can easily by evaluated
numerically.   Results are shown in figure~(\ref{fig3}) for
the five possible mixed spin tetrahedra.  
For purposes of comparison, all different exchange couplings 
have been set equal to the single value $J$.  

The entropy increases from a lower bound set 
by the groundstate degeneracy (which is greatest
for the tetrahedra with an odd number of spin $S=1$
moments, for which the ground state has a net spin of 1/2)
to and upper bound set by the total number of
spin degrees of freedom for each tetrahedron
(which is greatest for the 
tetrahedron with four spins $S=1$).
The curves for the entropy of the different
tetrahedra therefore cross.
This crossover from collective ground state to individual 
spin degrees of freedom takes place on a scale of temperatures
of order of the exchange coupling constant $J$, and the 
entropy has a point of inflection for $T \sim J/2$.
A more realistic parameterization of the 
exchange constants $\{J_1,J_2,J_3\} \ne J$
modifies the details of the crossover 
but does not affect the high or low 
temperature limits.

Similarly, we can evaluate the specific heat of the system
\be
c_{V} &=& 
   \frac{\langle E^2 \rangle - \langle E \rangle^2}{T^2}
\en
in terms of its the mean square energy
\be
\langle E^2 \rangle &=& \frac{1}{Z} \sum_n E_n^2 e^{-\frac{E_n}{T}} 
\en
Results are shown for the same set of tetrahedra in 
figure~(\ref{fig4}).   Once again, for purposes
of comparison, all exchange constants have been set 
equal to $J$.

The heat capacity of the tetrahedra at temperatures $T \ll J$
vanishes since the first excitation energy of the tetrahedron
occurs at finite energy $E_1 \sim J$.  The heat capacity is peaked
for $T \sim J/2$, where the entropy has its point of inflection,
and tends to zero at high temperatures as the entropy of the 
individual spins in the tetrahedron are saturated.   For temperatures
$T > J$, where individual spins predominate, the heat capacity is greatest 
for the tetrahedron with four large spins, since it has the greatest number of
degrees of freedom.

\subsubsection{Magnetic susceptibility}

\begin{figure}[tb]
\begin{center}
\leavevmode
\epsfxsize \columnwidth
\epsffile{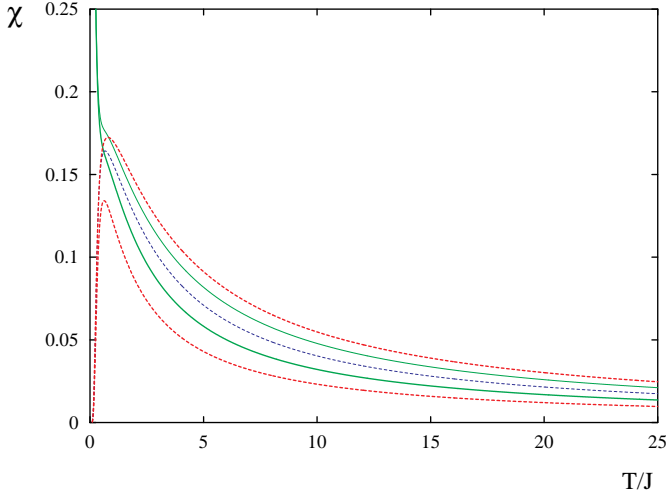}
\caption{
Magnetic susceptibility of spin $1/2$, spin $1$ and mixed
spin tetrahedra as a function of temperature.
From top to bottom at RHS plot, 
tetrahedra with --- 
4 spin 1 (dotted line),
3 spin 1 and 1 spin 1/2 (solid line), 
2 spin 1 and 2 spin 1/2 (dotted line),
1 spin 1 and 3 spin 1/2 (solid line),
4 spin 1/2 (dotted line).
All couplings set equal to J.}
\label{fig5}
\end{center}
\end{figure}

\begin{figure}[tb]
\begin{center}
\leavevmode
\epsfxsize \columnwidth
\epsffile{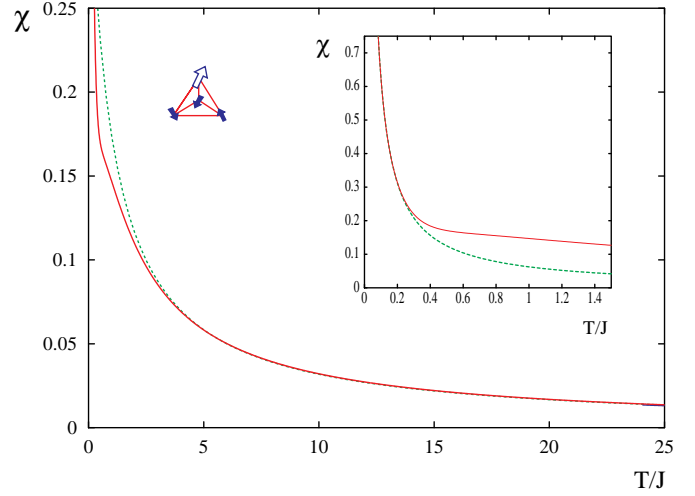}
\caption{
Magnetic susceptibility of isolated tetrahedron 
with one spin 1 and 3 spin 1/2 showing crossover
between different Curie laws at high and low 
temperatures (inset). 
}
\label{fig7}
\end{center}
\end{figure}

The magnetization of the tetrahedron in the presence of a magnetic 
field is given by
\be
\label{eqn:mag}
M &=& \frac{1}{Z} \sum_{\Omega\sigma\Sigma} g(\Omega,\sigma,\Sigma)
   \exp\left(-\frac{E(\Omega,\sigma,\Sigma)}{T}\right) 
      \no\\ && \quad \times 
   \Omega F_{\Omega}\left(\frac{h\Omega}{T}\right)
   B_{\Omega}\left(\frac{h\Omega}{T}\right)
\en
where $ B_{\Omega}(x)$ is the Brillouin function
\be
B_{\Omega}\left(x\right) 
   &=& \frac{(2\Omega+1)}{2\Omega}
   \coth \left(\frac{(2\Omega+1)x}{2\Omega}\right)\no\\
   && \qquad
   - \frac{1}{2\Omega} 
     \coth \left(\frac{x}{2\Omega}\right)
\en
We define the susceptibility {\it per site}
of the tetrahedron by 
\be
\chi^{\mc Tet} (T) 
   &=& \frac{1}{4} \frac{\partial M}{\partial h}
   \approx \frac{1}{4} \frac{M}{h}
\en
which in the limit of small $h/T$, gives
\be
\chi^{\mc Tet} (T)
   &=& \frac{1}{12T} \frac{1}{Z}
     \sum_{\Omega\sigma\Sigma} g(\Omega,\sigma,\Sigma) 
      \no\\ && \quad \times 
     \Omega(\Omega + 1) (2\Omega + 1)
      \no\\ && \qquad \times 
     \exp\left(-\frac{E(\Omega,\sigma,\Sigma)}{T}\right) 
\en
where, to the same level of approximation
\be
Z &\approx& 
   \sum_{\Omega\sigma\Sigma} g(\Omega,\sigma,\Sigma) (2\Omega + 1)
      \no\\ && \qquad \times 
    \exp\left(-\frac{E(\Omega,\sigma,\Sigma)}{T}\right) 
\en
Results for the susceptibility of the five different tetrahedra 
are shown in figure~(\ref{fig5}).
Further details of two mixed spin tetrahedra are shown in 
figures~(\ref{fig7})~and~(\ref{fig11}).
Once again, in order to simplify comparisons, all exchange 
constants have been set equal to $J$.

In the limit where $T/J \to \infty$ we must recover a Curie--Weiss
susceptibility
\be
\chi^{\mc Tet} (T\to \infty) \to \frac{C}{T + \theta}
\en
where the coefficient $C$ represents the contribution of 
an individual spin to the susceptibility and $\theta$
is the Curie temperature associated with interactions
between spins within the same tetrahedron.
In practice the crossover to this high temperature regime
occurs for $T \sim 5J$.

The value of $C$ is given by $C_S = S(S+1)/3$
only when the tetrahedral subsystem consists 
entirely of spin $S$ local moments.  For the mixed spin case,
it is an average of the different $C_S$'s of the 
different spins within the tetrahedron.  In general
it can be written as  
\be
C &=& \frac{1}{12} \frac{N_1}{N_0} 
\en
where 
\be 
N_0 &=& \sum_{\Omega\sigma\Sigma} g(\Omega,\sigma,\Sigma)
                    (2\Omega + 1)
\en
is the total number of states of the system and
\be 
N_1 &=& \sum_{\Omega\sigma\Sigma} g(\Omega,\sigma,\Sigma)
                    (2\Omega + 1) \Omega(\Omega+1)
\en
is a number determined by the degeneracy $g(\Omega,\sigma,\Sigma)$
of the states of the mixed spin tetrahedron.  

Similarly, for a tetrahedron with a single size of spin, the 
Curie temperature $\theta$ associated with interaction between spins 
can be written  $\theta_S = z_0 J S(S+1)$ where $z_0 = 3$ 
is the number of neighbouring spins within the same tetrahedron.
In the mixed spin case, this generalizes to 
\be 
\theta 
   &=& \frac{J_{\Omega}}{2} \left(\frac{N_2}{N_1} - \frac{N_1}{N_0}\right)
   \no\\ && \quad
     + \frac{J_{\sigma}}{2} \left(\frac{N_2^{\sigma}}{N_1} 
                             - \frac{N_1^{\sigma}}{N_0} 
                          \right)
   \no\\ && \qquad
     + \frac{J_{\Sigma}}{2} \left(\frac{N_2^{\Sigma}}{N_1} 
                             - \frac{N_1^{\Sigma}}{N_0}
                          \right)
\en
where the various numerical factors are given by
\be
N_1^{\sigma} &=& \sum_{\Omega\sigma\Sigma} g(\Omega,\sigma,\Sigma)
                      (2\Omega + 1) \sigma(\sigma+1)\\
N_1^{\Sigma} &=& \sum_{\Omega\sigma\Sigma} g(\Omega,\sigma,\Sigma)
                      (2\Omega + 1) \Sigma(\Sigma+1)\\
N_2 &=& \sum_{\Omega\sigma\Sigma} g(\Omega,\sigma,\Sigma)
                      (2\Omega + 1) \Omega^2(\Omega+1)^2\\
N_2^{\sigma} &=& \sum_{\Omega\sigma\Sigma} g(\Omega,\sigma,\Sigma)
                      (2\Omega + 1)  \Omega(\Omega+1)
   \no\\ && \qquad \times
      \sigma(\sigma+1)\\
N_2^{\Sigma} &=& \sum_{\Omega\sigma\Sigma} g(\Omega,\sigma,\Sigma)
                      (2\Omega + 1) \Omega(\Omega+1)
   \no\\ && \qquad \times
      \Sigma(\Sigma+1)
\en
Values of the coefficient $C$ and the Curie temperature $\theta$ for
different mixed spin tetrahedra are given in table~(\ref{table1a}). 
As a compact notation we refer to a tetrahedron with one spin one
and three spin half moments as $(1,1/2,1/2,1/2)$, etc.
The related numerical coefficients, and degeneracy factors are
listed in an Appendix.
Mean field corrections to $\theta$ 
will be discussed below.

\begin{table*}[tb]
\caption{
Curie coefficients and temperatures for 
tetrahedra with different mixtures of spin.
Mean field corrections to the Curie
temperature assume $z_{eff}=3$.
}
\label{table1a}
\begin{tabular}{llll} 
\hline\noalign{\smallskip}
      & C  & $ \theta$  & $\Delta\theta^{\mc MF}$ \\  
\noalign{\smallskip}\hline\noalign{\smallskip}
($\half,\half,\half,\half$) & 0.25 & 
   $0.75 J_1$ & $0.75 J_{eff}$\\
\noalign{\smallskip}\hline\noalign{\smallskip}
($1,\half,\half,\half$) & 0.351 & 
   $0.159 J_1 + 0.855J_3$ & 
   \\
\noalign{\smallskip}\hline\noalign{\smallskip}
($1,1,\half,\half$) & 0.458 & 
   $0.0068J_1 + 0.485J_2 + 0.727J_3$ & 
   $1.374 J_{eff}$\\
\noalign{\smallskip}\hline\noalign{\smallskip}
($1,1,1,\half$) & 0.558& 
   $0.759 J_2 + 0.923J_3$ & 
   \\
\noalign{\smallskip}\hline\noalign{\smallskip}
($1,1,1,1$) & 0.667 & 
   $2.00 J_2$ & $2.00 J_{eff}$\\
\noalign{\smallskip}\hline
\end{tabular}
\end{table*}

At low temperatures $T \ll J$ the behaviour of the susceptibility 
depends on the spin of the ground state of the tetrahedron.
The tetrahedra with an even number of spin $S=1$ moments
have singlet ground states, with exponentially activated
magnetic susceptibility (see figure~(\ref{fig11})).
At intermediate temperatures $T \sim J$ the susceptibility 
for these systems is strongly peaked.
The tetrahedra with an odd number of spin $S=1$ moments
(see figure~(\ref{fig7})) 
have a susceptibility diverging as $1/4 \times 3/4T$ for $T \to 0$.
At intermediate temperatures the susceptibilities of these tetrahedra 
cross over smoothly to the high temperature Curie--Weiss law.

\subsection{Mean Field Theory}

\begin{figure}[tb]
\begin{center}
\leavevmode
\epsfxsize \columnwidth
\epsffile{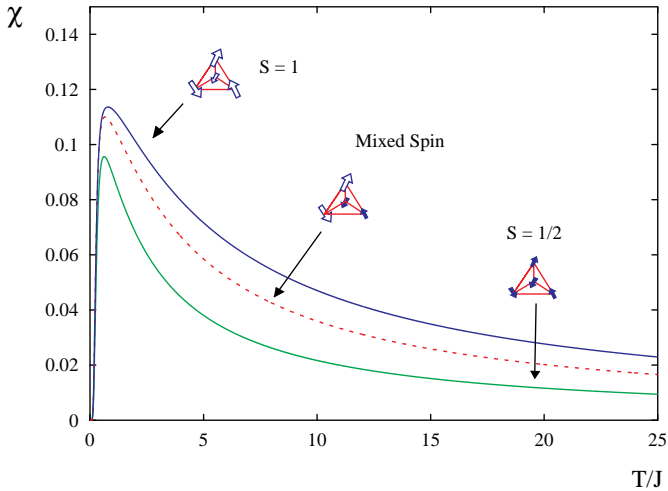}
\caption{
Magnetic susceptibility of spin $1/2$, spin $1$ and mixed
spin tetrahedron with two spin $1$ moments 
as a function of temperature, including mean
field interactions between tetrahedra.
All couplings set equal to J.}
\label{fig11}
\end{center}
\end{figure}

\begin{figure}[tb]
\begin{center}
\leavevmode
\epsfxsize \columnwidth
\epsffile{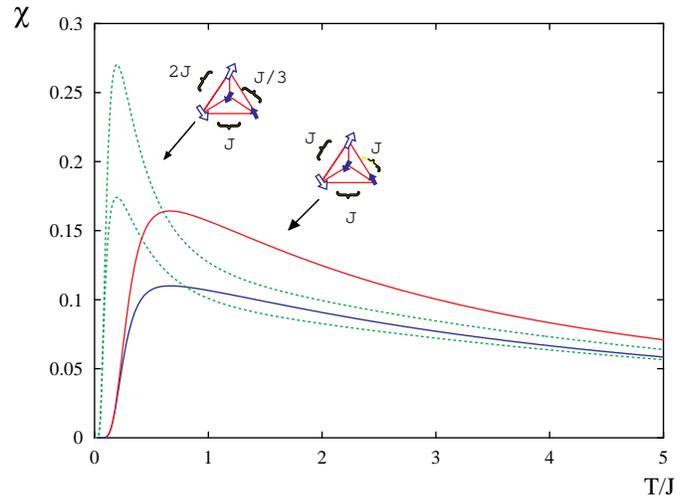}
\caption{
Magnetic susceptibility of isolated tetrahedron with two spin 1
and two spin 1/2 (upper pair lines) and mean field susceptibility
of equivalent lattice model (lower pair lines).
Solid lines are for \mbox{$J_1 = J_2  = J_3 = J_{eff} = J$}.
Dashed lines are for \mbox{$J_3  = J$}, J$_1$ = J/3,
J$_2$=2J, J$_{eff}$ = 0.680441, chosen so that the meanfield
Curie temperature is the same in each case.
}
\label{fig10}
\end{center}
\end{figure}

As suggested by  Garci\'a--Adena and Huber \cite{huber}, 
we can construct a mean field 
theory for the Heisenberg model on a pyrochlore lattice
by considering each spin within a tetrahedron on the $A$ sublattice 
to feel only the average effect of interactions with spins 
in other tetrahedra.  Where the groundstate of 
each tetrahedron is assumed to be a spin singlet, for example
in the three integer total spin cases considered above,
the different tetrahedra interact with one another only when a magnetic
field is applied.  In this case, the effective field felt by any given 
spin is {\it reduced} by its AF interaction with the induced 
magnetization of neighbouring tetrahedra, and the susceptibility 
of the system is accordingly modified to 
\be
\chi^{\mc MF}(T) &=& \frac{\chi^{\mc Tet}(T)}{
          1 + z_{eff}J_{eff}\chi^{\mc Tet}(T)}
\en
where $z_{eff}$ is the number of neighbouring spins
in {\it different} tetrahedra, and $J_{eff}$ is the effective
exchange interaction for the ``missing'' bonds of the B sublattice.
In theory, for a single spin system with a single type of spin and 
only nearest neighbour interactions
$z_{eff}=z_0=3$ and $J_{eff}=J$.  But in practice, even for systems
with only one type of spin, when it comes to comparison with
experiment, the product $z_{eff}J_{eff}$ is probably 
best regarded as an adjustable parameter \cite{huber}.

The new mean field Curie temperature is related to the Curie
temperature of an isolated tetrahedron by 
\be
\theta_{\mc MF} &=& \theta + z_{eff}J_{eff}C
\en 
The coefficient $C$ of the high temperature susceptibility is
of course independent of interaction and so unchanged.

For simplicity, we have limited our discussion here to the case of 
tetrahedra with singlet groundstates, where the generalization
of the theory presented by \cite{huber} is most straightforward.
We note, however, that the tetragonal mean field 
theory can also be generalized to the tetrahedra with a net
spin in the groundstate, by assuming that tetrahedra on the 
A and B sublattices order anti--parallel to one another.
This leads to an additional crossover at low temperatures when
the N\'eel order of the tetrahedra melts, which will be discussed
elsewhere.

Results for the mean field susceptibilities of different tetrahedra 
are shown in figure~(\ref{fig11}).
For AF exchange interactions as defined above, the 
mean field corrections lead to an overall suppression
of the susceptibility, which is reflected in the increase of the 
Curie temperature calculated above.
The tetrahedra with an even number of spin $S=1$ moments
still show a peak in their susceptibility at $T \sim J$,
but this is now a less pronounced maximum.

\begin{figure}[tb]
\begin{center}
\leavevmode
\epsfxsize \columnwidth
\epsffile{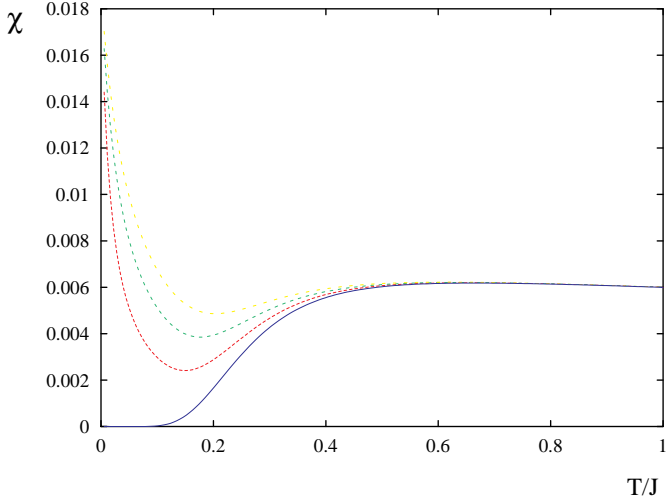}
\caption{
Mean field theory including different types of
mixed spin tetrahedra.  From top to bottom --- 
$\alpha$=0.3 (dotted line),
$\alpha$=0.2 (dotted line),
$\alpha$=0.1 (dotted line),
$\alpha$=0.0 (solid line),
where $\alpha$ is defined by equation~(\ref{eqn:alpha}). 
}
\label{fig:alpha}
\end{center}
\end{figure}

In the examples above we have set all the 
exchange constants $\{J_1,J_2,J_3,J_{eff}\} = J$.
In figure~(\ref{fig10}) we illustrate the effect
of relaxing this constraint on the magnetic susceptibility 
of an individual tetrahedron with two spin $S=1$ moments, 
and on the mean field theory for  a lattice of such
tetrahedra.  Lowering the coupling between the two
spin $S=1/2$ moments to $J_1 = J/3$ while increasing that
between the spin $S=1$ moments to $J_2 = 2J$ leads to a 
sharper peak in the susceptibility at lower temperatures, 
as more excitations become accessible at low temperatures.

However, since the high temperature susceptibility is of Curie law 
form in either case, these modifications are pronounced
only on a scale of $T \sim J \to 2J$.  From table~(\ref{table1a})
we see that while the low energy scale $J_1$ is important for the 
low temperature susceptibility, the Curie temperature of the 
tetrahedron is extremely insensitive to change in $J_1$.
In the example plotted,
the mean field coupling $J_{eff}$ has been adjusted so as to
compensate for the new values of $J_1$ and $J_3$, giving 
the same Curie temperature and therefore the same
high temperature susceptibility.  In practice the two 
models become indistinguishable for $T > 5J$.
This means that a representative average ``$J$'' can 
be extracted from knowledge of the high temperature
susceptibility of a system described by this model.

It is also interesting to consider the case of a lattice
of such tetrahedra which is modified by the 
inclusion of a low density of ``impurity''
tetrahedra with greater (or lesser) total spin.
To make this concrete, let us suppose that on average each 
tetrahedron contains two 
spin $S=1$ and two spin $S=1/2$ moments, but that in some 
fraction $\alpha/2$ of tetrahedra, there are in fact
three spin $S=1$ moments, and in an equal number three
spin $S=1/2$ moments.
Then the mean field susceptibility
is modified to 
\be
\label{eqn:alpha}
&&\chi^{\mc Tet}(\alpha,T) 
   = \alpha \chi^{\mc Tet}_{(11\half\half)}(T)
   \no\\ 
&& \quad + \frac{\alpha}{2} \left[  
         \chi^{\mc Tet}_{(111\half)}(T)
         + \chi^{\mc Tet}_{(1\half\half\half)}(T)
    \right]
\en
Results are shown for $J_1=J_2=J_3=J_{eff}=J$ and a range of values 
of $\alpha$ in figure~(\ref{fig:alpha}).   At high temperatures
the system must show its ``average'' character in a well defined
Curie law, and the redistribution of moments between different
tetrahedra is irrelevant.   In fact we do not even need to consider 
the temperatures $T > 5J$ for which the Curie law is valid --- for 
temperatures $T > J/2$ the increased susceptibility of the tetrahedra
with three spin $S=1$ moments exactly cancels the reduced susceptibility
of the tetrahedra with three spin $S=1/2$ moments and all results
collapse onto the curve for $\alpha = 0$.  At lower temperatures
the presence of the tetrahedra with a net groundstate spin leads to 
an upturn in the susceptibility.  This becomes steadily more pronounced 
as $ \alpha \to 1$, although the mean field theory cannot be relied
upon in this limit.  

Further generalizations of the meanfield theory introduced
in \cite{huber} for geometrically frustrated magnets with a 
single type of spin have been given in \cite{angel4,angel5,angel6}.

\section{Magnetic susceptibility of LiV$_2$O$_4$}
\label{experiment}

LiV$_2$O$_4$ is the first $d$--electron system to exhibit true 
``heavy Fermion'', \ie Pauli paramagnetism and 
(approximately) linear specific heat at low temperature, 
both with strongly enhanced coefficients, but with 
a Wilson ratio W$\sim$1.7 \cite{LiV2O4wilson}).  
The Fermi energy lies in the vanadium t$_{2g}$ $d$ electron 
bands, which are in total half filled,
giving an average of $1.5$ $d$ electrons per vanadium lattice 
site.
A presumed strong Hund's rule coupling implies that 
in an atomic basis, each site possesses either a spin 
S=1/2 or spin S=1 moment, according to whether one or 
two d electrons are found on that site.

At very low temperatures
the resistivity of LiV$_2$O$_4$ increases as $T^2$
\cite{LiV2O4resistivity}, and the 
Nuclear relaxation rate obeys the Korringa law $1/T_1T = const$,
as would be expected of a Fermi liquid with well defined 
quasi--particles \cite{LiV2O4nmr}.   However this behaviour 
breaks down at about 4~K, and LiV$_2$O$_4$ is a poor conductor,
with low temperature resistivity intermediate between that of a good metal 
(Ag, Cu) and that of an intrinsic semi--conductor (Si,Ga).
In addition, the entropy associated with the low
energy electronic excitations of the system is very large.  An 
estimate made by integrating the heat capacity (after appropriate
background subtractions for phonon and impurity contributions)
gives approximately 
$0.5 k_B\log 2$ per site at 50~K \cite{LiV2O4resistivity}.  
This value is much greater
than that for any other $d$--electron system, and should be compared with 
the maximum entropy per site of $k_B \log 2$ 
for a single spin S=1/2 degree of freedom.   
If we interpret the low temperature heavy Fermion behaviour 
of LiV$_2$O$_4$ naively in terms of an enhanced mass, electronic 
quasi--particles are approximately as massive as muons.   
Bandstructure calculations, on the other hand, suggest a relatively small
mass correction and underestimate the specific coefficient 
$\gamma$ by a factor of 25 \cite{LiV2O4lda,LiV2O4twoelectron}.

At about $20K$, the electronic physics of LiV$_2$O$_4$ undergoes 
marked change, visible in measurements of resistivity, heat capacity, 
susceptibility and Hall coefficient \cite{LiV2O4resistivity}.  
This crossover has sometimes been identified 
with the coherence temperature for an s--f heavy Fermion system, 
and a number of authors have suggested a minimal Kondo lattice model 
for LiV$_2$O$_4$ in which two--thirds of the 
d--electrons play the role of local moments (a single 
spin S=1/2 per site), and the remaining third are itinerant.
Various mechanisms have been proposed to justify 
treating subsets of the vanadium t$_{2g}$ $d$ electrons on a different 
footing, and hints of local moment physics for d electrons 
are even seen in some band structure calculations \cite{LiV2O4twoelectron},
but no real sign of Kondo physics (\eg logerythmic corrections to
resistivity) are seen immediately above the ``transition'' at 20~K. 

The magnetic susceptibility of LiV$_2$O$_4$ displays a
number of interesting features over a wide range of temperatures.  
At low temperatures (T $<$ 40~K) it exhibits a weakly temperature 
dependent Pauli paramagnetic susceptibility, but with a massively
enhanced value of $\chi \sim5 \times 10^{-3}$ per mole 
vanadium.
This crosses over smoothly to what has generally been 
interpreted as Curie law behaviour, but with different coefficients in 
different temperature ranges 100--500~K and 
500--1000~K \cite{muhtar,hiyakawa}.
Over the same wide range of temperatures the resistivity continues 
to increase slowly but monotonically, and comfortably exceeds the Mott--Regel 
limit \cite{LiV2O4resistivity}.

In what follows we will make the approximation of treating 
LiV$_2$O$_4$ as an insulating 
Heisenberg system of magnetic moments on a pyrochlore lattice.
This is not unreasonable, as the magnetic susceptibility of LiV$_2$O$_4$
varies on a scale typical of Heisenberg exchange integrals 
(10--100K), and not on the scale of the Hund's rule coupling 
or d--electron bandwidths found from LDA calculations 
(both $\sim 10^4$K), 
and because the naive mean free path for electrons is 
of atomic proportions.
Furthermore, the frustrated geometry of the pyrochlore 
lattice means
that spin coherence lengths will also be small, so the 
tetrahedral mean field theory developed above
can be expected to provide a reasonable starting point
for discussing its magnetic susceptibility.
In what follows we will consider three different scenarios
for the magnetic physics of LiV$_2$O$_4$, using our simple
model and the experimentally measured susceptibility to
place constraints on each.

The theoretical predictions for magnetic susceptibility per spin 
given above can be related to the experimentally measured susceptibility
in emu per mole vanadium (equivalently cm$^3$~[mol~V]$^{-1}$)
according to
\be 
\chi^{exp}(T) &=& 0.375 g_L^2 \chi^{theory}(T)
\en
where $g_L \sim 2.0$ is the Land\'e g--factor for the 
coupling of a magnetic field to the spin of a vanadium ion.  
We note that experimental susceptibilities are 
often quoted in \mbox{emu mol$^{-1}$}, \ie per {it mole--formula--unit}.
One mole of LiV$_2$O$_4$ contains two vanadium ions.

\begin{table*}[tb]
\caption{
Exchange coefficients, exponents, and Land\'e g--factors 
found from fits to data.
}
\label{table1b}
\begin{tabular}{llll} 
\noalign{\smallskip}\hline
\hline\noalign{\smallskip}
      &  Kondo \etal \cite{LiV2O4chi} 
      & Muhtar \etal \cite{muhtar}  
      & Hayakawa \etal \cite{hiyakawa}    \\  
\noalign{\smallskip}\hline
\noalign{\smallskip}\hline
First Scenario : & & & \\
\noalign{\smallskip}\hline
 $g_{L}$  & $1.6$  & -- & --\\
\noalign{\smallskip}\hline
 $J$ & 17.8 K   & -- & --\\
\noalign{\smallskip}\hline
\noalign{\smallskip}\hline
Second Scenario - high temperature regime :& & & \\
\noalign{\smallskip}\hline
 $g_{L}$  & -- & $2.04$ & $2.14$ \\
\noalign{\smallskip}\hline
 $J$ & -- & 119 K & 122 K\\
\noalign{\smallskip}\hline
\noalign{\smallskip}\hline
Second Scenario - low temperature regime :& & & \\
\noalign{\smallskip}\hline
 $\chi_0$ 
  & $0.26 \times 10^{-3}$ emu/mol V 
  & $0.18 \times 10^{-3}$ emu/mol V 
  & $0.12 \times 10^{-3}$ emu/mol V\\
\noalign{\smallskip}\hline
 $J$ & 26 K& 25 K & 13 K\\
\noalign{\smallskip}\hline
\noalign{\smallskip}\hline
Third Scenario :& & & \\
\noalign{\smallskip}\hline
$\alpha$ & -- & 0.74 & 0.80\\
$T_0$ & -- &0.021 & 0.053\\
\noalign{\smallskip}\hline
\noalign{\smallskip}\hline
\end{tabular}
\end{table*}

\subsection{First scenario - mixed valent local moments near to charge order.}

\begin{figure*}[tb]
\begin{center}
\leavevmode
\epsfysize = 100.0mm
\epsffile{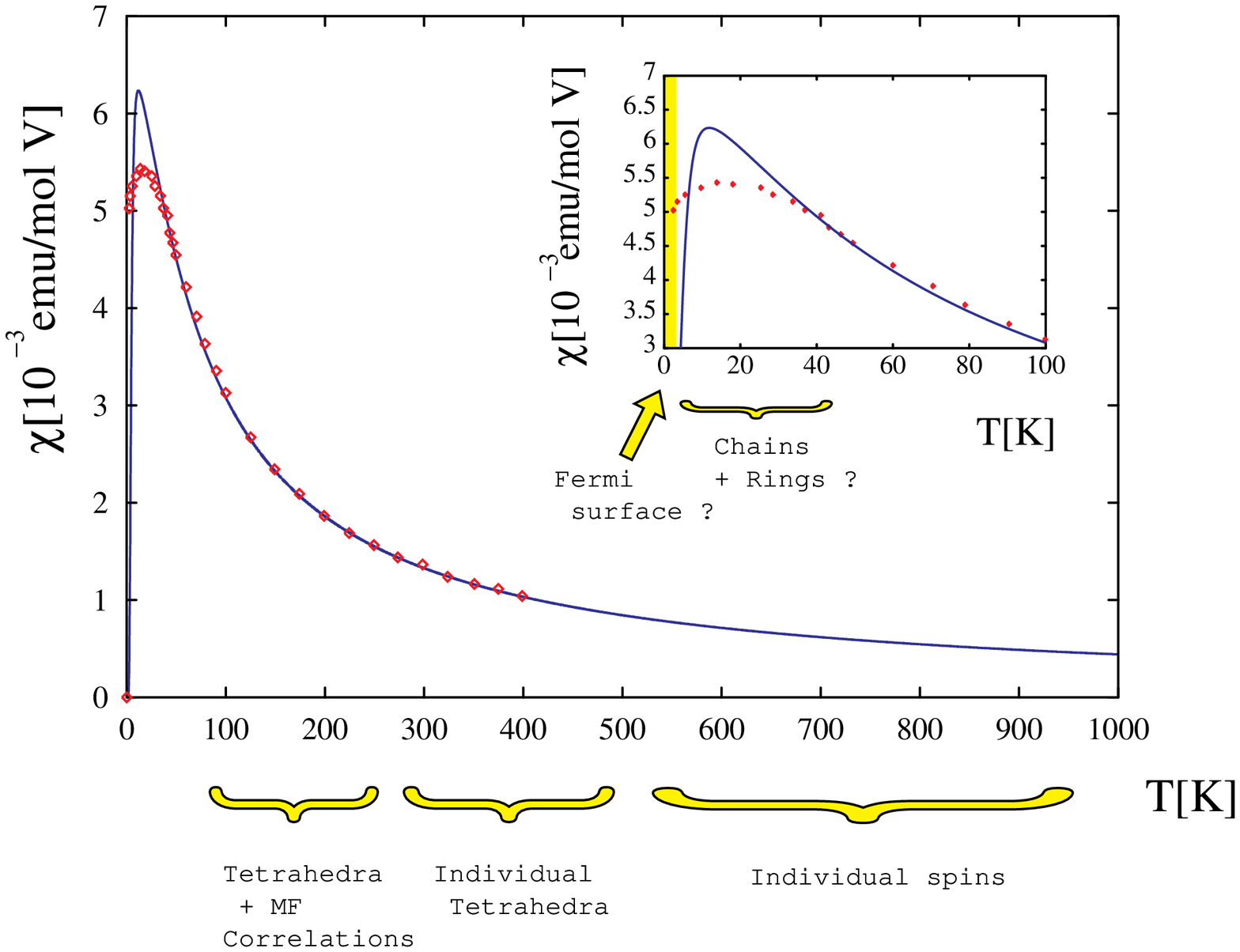}
\caption{First scenario ---
fit of mixed--spin tetragonal mean field theory to 
the experimentally measured magnetic susceptibility of 
LiV$_2$O$_4$ over the temperature range 0--400K, taken 
from \cite{LiV2O4chi},
using the two adjustable parameters 
\mbox{$g=1.6$} and \mbox{$J=17.8$ K}.  Annotation on the 
temperature axis shows the type of correlation between spins.  
Inset --- how the fit breaks down at low temperatures. 
A possible interpretation of the electronic state of the system in 
terms of charge order correlations in different low temperature 
ranges is given.}
\label{scenarioI}
\end{center}
\end{figure*}

The vanadium atoms in LiV$_2$O$_4$ occur in two valence states, 
d$^1$ (V$^{4+}$) and d$^2$ (V$^{3+}$).
Both of these have an incomplete shell of d--electrons, and 
vanadium has a strong Hund's first rule coupling, so both 
have a net magnetic moment --- $S=1/2$ in the case of V$^{4+}$, 
and $S=1$ in the case of V$^{3+}$.

Another important fact is that LiV$_2$O$_4$ is close to charge 
order.  This could be anticipated by analogy with other mixed valent 
transition metal spinels --- for example those 
Ferrites which
undergo a Verwey (charge ordering) transition.
To explain this, Anderson invoked a ``tetrahedron rule'' requiring
that charge balance be satisfied within each tetrahedron, \ie, 
that each tetrahedron should have two of the high, and two of the low
ionization states \cite{anderson}.  If all events violating the 
tetrahedron rule are neglected, the resulting state is a
charge ordered magnetic insulator, with dynamics determined by 
the residual (antiferromagnetic) Heisenberg exchange integrals.

While LDA estimates suggest that the energy associated with
Coulomb interaction between V$^{3+}$ and V$^{4+}$ ions on
neighbouring sites is lower than the threshold for charge
order \cite{yaresko},
experimentally LiV$_2$O$_4$ {\it does} charge order under 
pressure \cite{chargeorder}.
It has therefore been suggested 
by Fulde \etal \cite{fulde} that the application of the tetrahedron 
rule to LiV$_2$O$_4$ provides a way of explaining its heavy Fermion 
behaviour at low temperatures.  

The essential ingredient of this theory is the emergence of 
one--dimensional correlations between spins as a result
of the tetrahedron rule.  Since each tetrahedron has two
spin one and two spin half moments, every vanadium atom must 
be connected to 
two vanadium atoms with same moment, 
one within its tetrahedron, and one in a neighbouring tetrahedron.
This means that the lattice of tetrahedra
can be divided into Heisenberg chains of spin half or spin one moments.
These chains may close to form rings, with a minimum length
of six spins.  The remaining simplification is that the 
interaction between neighbouring chains is neglected,
so that the spin one chains have (Haldane--) 
gapped excitations, while the spin half chains 
have low lying fermionic excitations with linear 
specific heat.
These, and not the dressed electronic quasiparticles
of the more familiar rare earth heavy Fermion compounds,
are the heavy Fermions of Fulde's theory.

Since this scenario for calculating the low
temperature susceptibility of LiV$_2$O$_4$
is based on the tetrahedron rule, and
treats LiV$_2$O$_4$ as an insulator, it is natural to
extend it to higher temperatures using the tetragonal
mean field for the theory described above.   Figure~\ref{scenarioI}
shows a mean field fit to the experimental susceptibility of LiV$_2$O$_4$
taken from \cite{LiV2O4chi},
based on tetrahedra with two spin one and two spin half moments.
The fit is excellent down to temperatures of order 20--30K, at which 
one might expect higher order correlation effects (for example 
the formation of chains and rings), and above all the fact 
that LiV$_2$O$_4$ is not an insulator, to become important.

Two adjustable parameters have been used for the fit, the Land\'e 
g--factor $g_{L}$ and a single representative Heisenberg exchange
integral $J=J_1=J_2=J_3=J_{eff}$.  The effective correlation number
$z_{eff}$ is set equal to three.
The parameters $g_{L}$ and $J$
are then uniquely determined by the Curie temperature 
$\theta = 47K $ and coefficient $C = 0.46 K emu/mol V$
which can be extracted 
from the Curie law behaviour of the susceptibility
on the range 100--400K.  The fit below 
100K then provides an independent test of the validity 
of the tetragonal mean field theory.
As can be seen in the inset to Figure~\ref{scenarioI}, the
tetragonal mean field theory appears to be very successful, at least
over the range of temperatures for which it can reasonably be 
applied.  

It is tempting to identify the temperature $T \sim J \approx 20K$
at which both the model and experimental susceptibiliies have
their maximum, and begin to diverge, as the scale for a crossover to 
a new low temperature state.  It is almost certainly true that the 
inclusion of processes which violated the tetrahedron rule, \ie the 
hopping of electrons between tetrahedra, would prevent the system
from achieving the singlet groundstate which the mean field theory predicts, 
and might reasonably lead to the emergence of a HF state.  

However, even above 40K, where the fit is very good, a number 
of important experimental facts remain  unaddressed by this scenario.  
One is that the Land\'e g--factor 
extracted from the fit is really too low, $g_L = 1.6$, as compared 
with the usual value of $g_L = 2.0$ found for bulk Vanadium.  
The tetrahedral mean field theory also has too great an entropy,
and therefore too great a heat capacity as compared with 
experimental estimates.   But the most challenging observation
is that published susceptibility data for temperatures 
of order 1000K appear 
to show a crossover to a different Curie law regime with 
$ C \approx 700 K emu/mol V$ and $\theta \approx 400 K$.  
It is this issue which we address in the following section.

\subsection{Second scenario - two different local moment regimes.}

\begin{figure*}[tb]
\begin{center}
\leavevmode
\epsfysize = 100.0mm
\epsffile{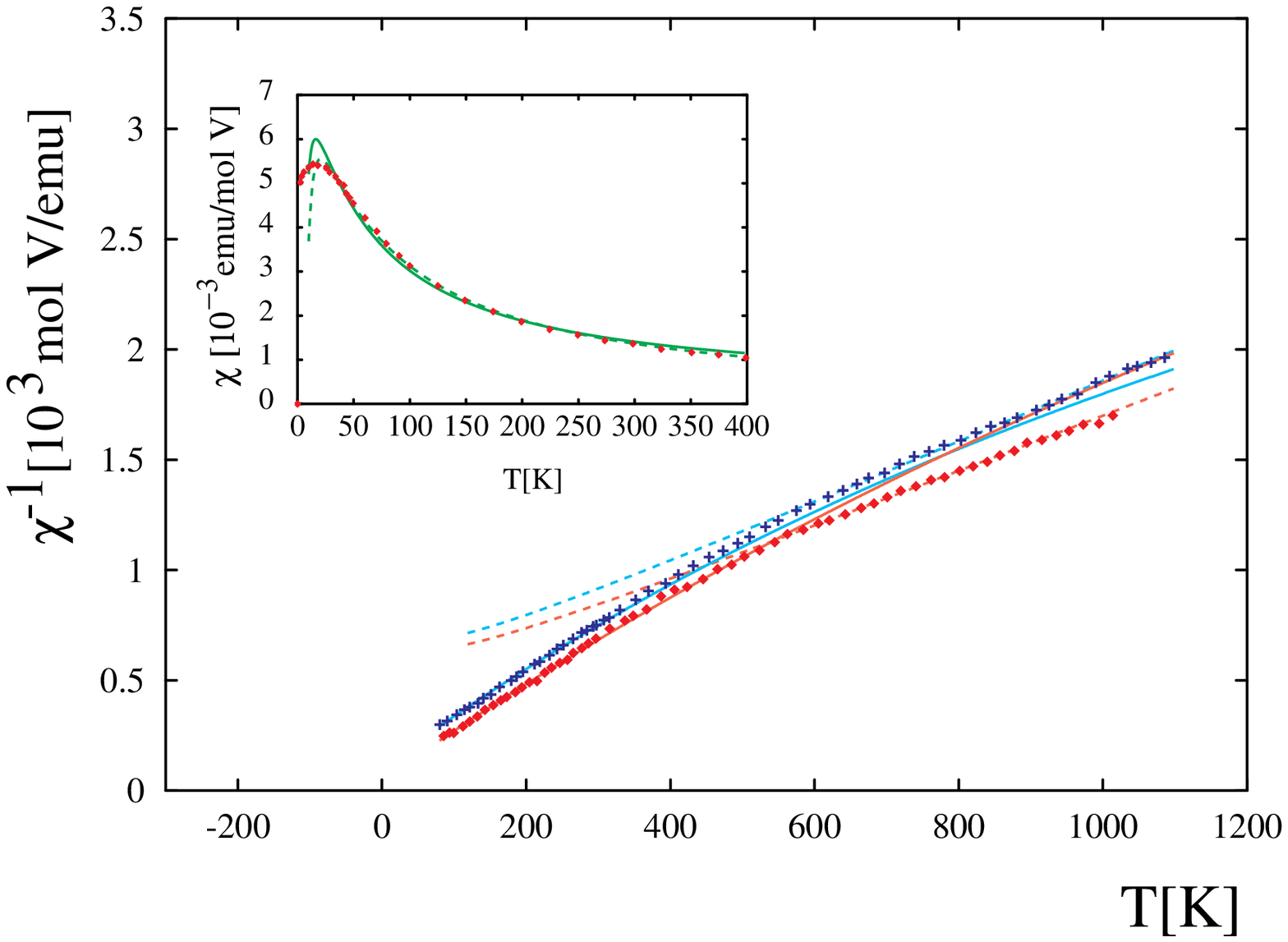}
\caption{Second scenario - inverse magnetic susceptibility of 
LiV$_2$O$_4$ over the range 0--1100K as quoted in 
\cite{muhtar} (diamonds),
\cite{hiyakawa} (squares), 
together with linear fits to the ``Curie Law'' 
behaviour seen at high (broken lines) and low temperatures
(unbroken lines). 
Inset --- low temperature fit to susceptibility 
taken from \cite{LiV2O4chi},
assuming a lattice of spin 1/2 moments and Land\'e g--factor g=2.04
(solid line), together with unconstrained fit (broken line).
}
\label{scenarioII}
\end{center}
\end{figure*}

Figure \ref{scenarioII} shows the {\it inverse} magnetic 
susceptibility of LiV$_2$O$_4$ between 100 and 1100 K, 
as reported by 
Muhtar \etal \cite{muhtar} and Hayakawa \etal \cite{hiyakawa}.
Above 600K, and below 400K Curie law behaviour 
is seen in the sense that the inverse susceptibility can be 
approximated by a
straight line with slope $1/C$ and intercept $\theta$.
However the values of the Curie temperature $\theta$ and
the the coefficient $C$ are quite different for the high and
low temperature regimes.   

In fact the value of the $C$ found at high temperatures
corresponds quite well to that which would be expected for 
an equal mixture of spin half and spin one moments (V$^{4+}$ and V$^{3+}$
ions), assuming a Land\'e factor $g_L = 2.0$, while the value
of $C$ found at low temperatures is much closer to that which would
be expected if each tetrahedral site had a spin half moment.

This has prompted the suggestion that the full spin moment of the 
$V$ atoms is seen only at high temperatures, while at lower
temperatures this moment is partially ``screened'' by correlations
between spins in such a way that only a net spin of one half
remains at each V site (see \eg \cite{piers}).  
The majority of theoretical attempts to explain heavy Fermion behaviour 
in LiV$_2$O$_4$ \cite{piers,rice,varma,lacroix,laad} take as
a starting point a tetrahedral lattice of spin half moments  
(often identified with the A$_{1g}$ representation of the 
V d--electron states), and assign the remaining half an electron
per site to an itinerant electron band (equivalently, E$_{g}$ states).
We will not attempt to review these theories here, but 
in the light of these models, 
it is clearly worth trying to obtain a ``self consistent'' fit 
to both the high and low temperature magnetic susceptibilities
of LiV$_2$O$_4$ within the overall scenario of two local moment
regimes.

Using the 
tetragonal mean field theory developed above, we therefore proceed as follows :
we first (least squares) fit the high temperature (600--1000K) susceptibility 
assuming an equal mixture of spin one and spin half moments, 
using the two adjustable parameters $J$ and $g_L$, as described
in the previous analysis.
We obtain values of $g_L=2.04$ and $J=119K$ for data taken from
Muhtar \etal \cite{muhtar} with a mean square error per point
of $\sigma = 0.0028$~emu/mol~V, as recorded in Table \ref{table1b}.  
Similarly, for data taken from Hayakawa \etal \cite{hiyakawa}
we obtain $g_L=2.14$, $J=122 K$ and $\sigma = 0.0024$~emu/mol~V
Then, using the 
value of the Land{\'e} g--factor $g_L$ obtained at high temperatures,
we fit the low temperature susceptibility (100--400K).
We do this assuming that each tetrahedral site has a localized spin half
moment, and that the contribution of the remaining itinerant
electrons can be lumped into a single paramagnetic constant
$\chi_0$ so that 
\be
\label{eqn:chi0}
\chi(T) &=& \chi_0 + \chi^{\mc MF} (T)
\en
where $\chi^{\mc MF} (T)$ is the mean field susceptibility of 
the lattice of spin half lattice tetrahedra \cite{note}.
As fit parameters we use $\chi_0$ and $J$, the exchange 
integral between the local spin half moments.
We obtain values 
$\chi_0=0.18 emu/mol V$ and $J=25 K$ with an error $\sigma = 0.030$
for data taken from
Muhtar \etal \cite{muhtar} and $\chi_0= 0.12$~emu/mol~V,
$J= 13 K$ and $\sigma = 0.047$~emu/mol~V
for data taken from Hayakawa \etal \cite{hiyakawa}.

In Figure \ref{scenarioII} we plot the data
described above, showing the high temperature
fit with dashed and the low temperature fit with unbroken
lines.  Each fit is good, within its own domain of validity.
The values of the fit parameters obtained from
data taken from \cite{muhtar} and \cite{hiyakawa}
are not quite the same, and the apparent variation
in values of the Land\'e g--factor $g_L$ might be 
a cause for concern.
However, as  
the fit parameters strongly depend on the 
how background contributions were subtracted from 
the experimental data, it is difficult to draw any strong 
conclusions about their precise values, and note
the absolute values of susceptibility quoted
for the heavy Fermion phase of LiV$_2$O$_4$
at very low temperatures also 
vary from group to group.

In the inset to Figure \ref{scenarioII} we show 
fits to the low temperature susceptibility of LiV$_2$O$_4$ 
as measured by \cite{LiV2O4chi} on the range 40--400K, using 
a model susceptibility of the form Eqn. \ref{eqn:chi0}.
As no high temperature data was available for this sample, 
we use both constrained (solid line)
and unconstrained (dashed line) values of $g_L$.
Values of the fit parameters are shown in Table \ref{table1b}.
For the constrained fit a value of $g_L = 2.04$ was taken 
from high temperature data for \cite{muhtar}, using which
values  of $\chi_0 = 0.26$~emu/mol~V and 
$J=26.0$~Kwere found, with a mean square 
error per point of $\sigma = 0.11$~emu/mol~V. 
The better fit was in fact obtained for the unconstrained
(three parameter) fit, for which $\chi_0=0$~emu/mol~V, $J=33 K$
and $g_L = 2.3$, and $\sigma = 0.052$~emu/mol~V.

To summarize, the assumption that LiV$_2$O$_4$ has two 
different local moment regimes as a function of temperature,
leads to fits to its magnetic susceptibility which
a) have physically parameters and b) have an error 
comparable to the uncertainty of the data.
However, by splitting the data into different temperature
regimes in this way we have not only assigned a physical 
meaning to the observed change in the slope of the inverse 
susceptibility, but also diminished what we learn from 
each fit --- almost any data set could be fitted piecewise, 
but cutting it into small enough pieces. 
Most importantly, our mean field theory can tell us nothing 
about how such a crossover between different local
moment regimes takes place, and this remains an important
question for microscopic theories of LiV$_2$O$_4$ to address.

In the section below we consider a radically different, and
admittedly speculative, way of understanding the magnetic susceptibility
of LiV$_2$O$_4$ over the temperature range 100--1100K, prompted
by experiments on Zn and Li doped samples.

\subsection{Third scenario - powerlaw scaling.}

\begin{figure*}[tb]
\begin{center}
\leavevmode
\epsfysize = 100.0mm
\epsffile{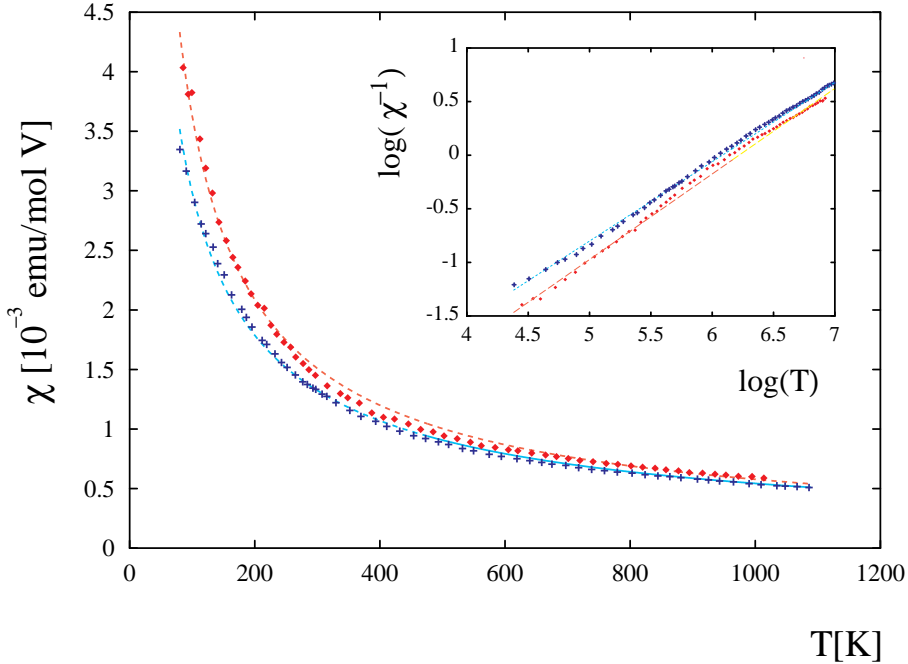}
\caption{
Third scenario --- power law fits to the magnetic susceptibility
of LiV$_2$O$_4$ as measured by \cite{muhtar} (diamonds) 
and  \cite{hiyakawa} (crosses), with exponents of 0.74 and 
0.80, respectively.
Inset --- inverse susceptibility and power law fits plotted 
on a log--log scale.
}
\label{scenarioIII}
\end{center}
\end{figure*}

LiV$_2$O$_4$ is by no means the only spinel oxide.
Many different systems have been synthesized, and 
all possess the same geometric frustration, which 
leads to a complex interplay of spin, charge 
and orbital order, which may in turn be linked
to lattice modes \cite{oleg}.
This is evident in the many different ground states
which these systems achieve --- some like ZnCr$_2$O$_4$
stabilizing spin order through a distortion of the lattice,
others, like AlV$_2$O$_4$, achieving charge order through
valence skipping.   
Crudely speaking, each material seeks
a means of reducing the high entropy associated with its 
frustrated geometry by playing off different competing
forms of order.
In LiV$_2$O$_4$ the low temperature state is a heavy 
Fermi liquid, and it seems reasonable to believe that
the system is a poor metal precisely because no 
one form of insulating order is achieved. 

It is possible to dope LiV$_2$O$_4$ by substituting 
the Zn for Li to give Li$_{1-x}$Zn$_x$V$_2$O$_4$ \cite{muhtar}, 
or by substituting Ti for V to give LiTi$_y$V$_{2-y}$O$_4$ \cite{hiyakawa}.
Zn is a magnetic impurity, and occupies the octahedral sites
in the spinel.
The inclusion of small 
concentrations of Zn forces LiV$_2$O$_4$ into a spin glass phase,
with a spin glass temperature which vanishes as
the number of Zn impurities tends to zero \cite{LiV2O4resistivity}.
The alternative ``parent'' compound Zn$_1$V$_2$O$_4$ is an 
AF Mott insulator.  
Ti is non--magnetic, and occupies tetrahedral sites in the spinel.
Small concentrations of Ti do not substantially alter the properties
of LiV$_2$O$_4$, but at larger dopings it undergoes a metal insulator
transition.
LiTi$_2$O$_4$ is a conventional superconductor with $T_c=13.7 K$.
The Curie coefficient $C$ extracted from the high temperature
susceptibilities of Li$_{1-x}$Zn$_x$V$_2$O$_4$ and LiTi$_y$V$_{2-y}$O$_4$
does appear to have the expected dependence on doping.

So LiV$_2$O$_4$ lies at a quantum critical point
for a transition into a spin glass
phase on doping, and is close to charge order
on the application of pressure.  It is therefore clear that there 
is quantum phase transition (probably, a line of critical points) 
close to the undoped, ambient pressure, ground state.  What influence,
if any, could this be expected to have ?  Quantum phase transitions
at zero temperature can manifest themselves at finite temperature
through the power law scaling of response functions.   Is there 
any evidence for scaling behaviour in LiV$_2$O$_4$ ?

We make the simple conjecture that the magnetic susceptibility
of LiV$_2$O$_4$ might be described by a simple power
law of the form 
\be
\label{eqn:powerlaw}
\chi(T) &=& A\left(\frac{T}{T_0}\right)^{\alpha}
\en
over a wide range of temperatures.

The magnetic susceptibility of LiV$_2$O$_4$ is shown plotted
on a log--log scale in the inset to Figure \ref{scenarioIII}.
If the temperature dependence of the data were a simple
power law, the data would lie on a straight line.  In the 
case of a Curie law, this straight line would have a gradient 
of one.   

A least squares fit to the susceptibility data reported by \cite{muhtar}
{\it over the full range of temperatures} (i.e. 100--1100K)
leads to an exponent $\alpha=0.74$, with a mean error per point
of $\sigma = 0.038$.  In the case of the data reported by 
\cite{hiyakawa}, fitting the full data set from 100--1000K,
we find an exponent $\alpha=0.80$, and an error per point 
$\sigma = 0.053$.   The errors of these fits are not as good
as those for the fits to Curie law behaviour at high temperatures,
but no worse than those for the self consistent low temperature 
fits within the two local moment scenario.  The fits are shown
on a linear scale in Figure \ref{scenarioIII}.

Existing evidence therefore does not rule out the possibility that,
rather than exhibiting a crossover between two different local
moment regimes, the magnetic susceptibility of LiV$_2$O$_4$
has a simple power law behaviour over a very wide range of 
temperatures.  Such powerlaw scaling would eventually have to 
``saturate'' in Curie law behaviour at very high temperatures,
when correlations between moments can legitimately be ignored.

\section{Conclusions}
\label{conclusions}

The magnetic susceptibility of the heavy Fermion spinel
LiV$_2$O$_4$ is puzzeling, not only in the size of the 
paramagentic contribution found at low temperatures, 
but in the way in which this crosses over to 
local moment behaviour at high temperatures.
We consider geomteric frustration to play an important
role LiV$_2$O$_4$ and have addressed this issue by extending 
a recently introduced tetragonal mean field theory for a Heisenberg
model on a pyrochlore lattice to allow for mixed valance.

Using this theory as a tool to make comparison with the experimentally
measured magnetic susceptibility of LiV$_2$O$_4$, we have considered a
number of different scenarios for the crossover from
(roughly) temperature indpendant paramagnetism below 40 K to
apparent Curie law behaviour at 1000K.  We find that fits based
on the tetragonal mean field thoery work well below 400K, for a 
wide range of parameter sets, suggesting that the geometry of the 
lattice plays an important role in determining the magnetic 
properties of LiV$_2$O$_4$.  However not all of these fits yield 
physically reasonable values of the Land\'e g--factor and so the 
low temperature
susceptibility alone cannot uniquely constrain the model used.

Considering the susceptibility from 100-1100K, we find that
fits based on the assumption of two different local moment 
regimes, and fits based on the ansatz of power law scaling, 
both provide a reasonable account of the data.  This leads
us to speculate that instability of LiV$_2$O$_4$ against
a spin glass state (on doping), and a charge ordered state
(under pressure) manifests itself a the non--analytic behaviour
of the magnetic susceptibility.

Further analysis of theory and experiment is needed to
distinguish between these scenarios.

{\it Acknowledgments.}

It is our pleasure to acknowledge the suggestions and 
encouragement of Peter Fulde, together with helpful
conversations with David Huber, Sinchiro Kondo, Hide Takagi 
and Victor Yushanghi.  We would also like thank Martin Albrecht 
for a careful reading of the manuscript.

This work was supported under the visitors program of MPI--PKS.

\appendix

\section{Degeneracy of a state with total spin $\Omega$ }
\label{appendix1}

The problem of how to find the degeneracy of a state with 
total spin $\Omega \leq NS$ of a system of $n$ spins of length $S$
was solved by Van Vleck \cite{vanvleck}.  Here we review his derivation,
which may then be generalized easily to a system of mixed spin.

We first consider the simpler problem of finding $g^z (M)$, the 
number of states of the system with $z$--component of total
spin $\Omega^z = M$.   By simple combinatorics, this is given by 
the coefficient of $x^M$ in the polynomial
\be
(x^S + x^{S-1} + \ldots + x^{-S})^n
\en
The first few terms in this series are easy to calculate and 
have obvious physical significance.  They also demonstrate the pattern
for finding further terms
\be
x^{nS} &\to& (x^S)^n \to 1 \\
x^{(n-1)S} &\to& (x^S)^{(n-1)}.1 \to n \\
x^{(n-1)S} &\to& (x^S)^{(n-2)}.1.1 + (x^S)^{(n-1)}.(x^{-S})^{(n-1)}
   \no\\
&\to& \frac{n!}{(n-2)!2!} + \frac{n!}{(n-1)!1!}
\en
If we consider instead a system of $n_1$ spins of size $S_1$ and 
$n_1$ spins of size $S_2$ the polynomial in question
becomes
\be
&&(x^{S_1} + x ^{S_1 - 1} + \ldots + x^{-S_1})^{n_1}
   \no\\&&\quad \times
(x^{S_2} + x ^{S_2 - 1} + \ldots + x^{-S_2})^{n_2}
\en

For the purposes of calculating the partition function of a
tetrahedron what we need is $g(\Omega)$, the number of possible 
states with total spin $\Omega$, and not $g^z(M)$, the number of states
with $\Omega^z = M$.  We find $g(\Omega)$ by setting up a difference
equation.  The number of possible states $g^z(M)$ with magnetization 
$M > 0$ must increase with decreasing $M$, since all states with total 
spin $\Omega \geq M$ contribute to $g^z(M)$.
It follows immediately that the required degeneracy 
$g(\Omega)$ is just the rate of change of $g^z(M)$ for $M=\Omega$, \ie 
\be
g(\Omega) &=& g^z(\Omega) - g^z(\Omega + 1)
\en 
This generalizes directly to the case of a mixed spin system.

\section{Degeneracies $g(\Omega,\sigma,\Sigma)$}
\label{appendix2}

\begin{table}
\caption{Degeneracy $g(\Omega)$ for states of 
  spin $S=1/2$ tetrahedron with total spin $\Omega$
  and values of associated coefficients.
}
\label{table1}
\begin{tabular}{llllll} 
\hline\noalign{\smallskip}
      & $\Omega=0$  & $\Omega=1$  & $\Omega= 2$\\  
\noalign{\smallskip}\hline\noalign{\smallskip}
S=1/2 & 2  & 3  &  1\\ 
\noalign{\smallskip}\hline\noalign{\smallskip}
\noalign{\smallskip}\hline\noalign{\smallskip}
$N_0$ & 16 & & \\
\noalign{\smallskip}\hline\noalign{\smallskip}
$N_1$ & 48 & & \\
\noalign{\smallskip}\hline\noalign{\smallskip}
$N_2$ & 216 & & \\
\noalign{\smallskip}\hline
\end{tabular}
\end{table} 

\begin{table}
\caption{
   Degeneracy $g(\Omega,\sigma)$
   for states of 
   tetrahedron with three \mbox{$S=1/2$} 
   and one \mbox{$S=1$} spins as 
   function of 
   total spin \mbox{$\Omega=\{1/2,3/2,5/2\}$}, 
   spin of \mbox{$S=1/2$} subsystem 
   \mbox{$\sigma = \{1/2,3/2\}$},
   and values of associated coefficients.   
   }
\label{table2}
\begin{tabular}{lll} 
   \hline\noalign{\smallskip}
    & $\sigma = 1/2$  & $\sigma =3/2$ \\
   \noalign{\smallskip}\hline\noalign{\smallskip}
 $\Omega = 1/2$   & 1  & 2 \\
   \noalign{\smallskip}\hline\noalign{\smallskip}
 $\Omega = 3/2$   & 1  & 2 \\
   \noalign{\smallskip}\hline\noalign{\smallskip}
 $\Omega = 5/2$   & 0  & 1 \\
   \noalign{\smallskip}\hline\noalign{\smallskip}
   \noalign{\smallskip}\hline\noalign{\smallskip}
$N_0$ & 24 &\\
\noalign{\smallskip}\hline\noalign{\smallskip}
$N_1$ & 101 &\\
\noalign{\smallskip}\hline\noalign{\smallskip}
$N_1^{\sigma}$ & 71 &\\
\noalign{\smallskip}\hline\noalign{\smallskip}
$N_2$ & 630 &\\
\noalign{\smallskip}\hline\noalign{\smallskip}
$N_2^{\sigma}$ & 331 &\\
\noalign{\smallskip}\hline
\end{tabular}
\end{table} 

\begin{table}
\caption{
   Degeneracy $g(\Omega,\sigma,\Sigma)$
   for states of 
   tetrahedron with two \mbox{$S=1/2$} 
   and two \mbox{$S=1$} spins as 
   function of 
   total spin \mbox{$\Omega=\{0,1,2,3\}$}, 
   spin of \mbox{$S=1/2$} subsystem 
   \mbox{$\sigma = \{0,1\}$}
   and spin of \mbox{$S=1$} subsystem 
   \mbox{$\Sigma=\{0,1,2\}$},
   and values of associated coefficients.
   }
\label{table3}
\begin{tabular}{lll} 
   \hline\noalign{\smallskip}
$\Omega=0$  & $\sigma = 0$  & $\sigma  =1$ \\
   \noalign{\smallskip}\hline\noalign{\smallskip}
$\Sigma=0$ &  1          &  0  \\
   \noalign{\smallskip}\hline\noalign{\smallskip}
$\Sigma=1$ &  0          &  1 \\ 
   \noalign{\smallskip}\hline\noalign{\smallskip}
$\Sigma=2$ &  0          &  0 \\
   \noalign{\smallskip}\hline\noalign{\smallskip}
   \noalign{\smallskip}\hline\noalign{\smallskip}
$\Omega=1$  & $\sigma = 0$  & $\sigma  =1$ \\
   \noalign{\smallskip}\hline\noalign{\smallskip}
$\Sigma=0$ &  0          &  1  \\
   \noalign{\smallskip}\hline\noalign{\smallskip}
$\Sigma=1$ &  1          &  1 \\ 
   \noalign{\smallskip}\hline\noalign{\smallskip}
$\Sigma=2$ &  0          &  1 \\
   \noalign{\smallskip}\hline\noalign{\smallskip}
   \noalign{\smallskip}\hline\noalign{\smallskip}
$\Omega=2$  & $\sigma = 0$  & $\sigma  =1$ \\
   \noalign{\smallskip}\hline\noalign{\smallskip}
$\Sigma=0$ &  0          &  0  \\
   \noalign{\smallskip}\hline\noalign{\smallskip}
$\Sigma=1$ &  0          &  1 \\ 
   \noalign{\smallskip}\hline\noalign{\smallskip}
$\Sigma=2$ &  1          &  1 \\
   \noalign{\smallskip}\hline\noalign{\smallskip}
   \noalign{\smallskip}\hline\noalign{\smallskip}
$\Omega=3$  & $\sigma = 0$  & $\sigma  =1$ \\
   \noalign{\smallskip}\hline\noalign{\smallskip}
$\Sigma=0$ &  0          &  0  \\
   \noalign{\smallskip}\hline\noalign{\smallskip}
$\Sigma=1$ &  0          &  0 \\ 
   \noalign{\smallskip}\hline\noalign{\smallskip}
$\Sigma=2$ &  0          &  1 \\
\noalign{\smallskip}\hline\noalign{\smallskip}
\noalign{\smallskip}\hline\noalign{\smallskip}
$N_0$ & 36 &\\
\noalign{\smallskip}\hline\noalign{\smallskip}
$N_1$ & 198 &\\
\noalign{\smallskip}\hline\noalign{\smallskip}
$N_1^{\sigma}$ & 54 &\\
\noalign{\smallskip}\hline\noalign{\smallskip}
$N_1^{\Sigma}$ & 144 &\\
\noalign{\smallskip}\hline\noalign{\smallskip}
$N_2$ & 1596 &\\
\noalign{\smallskip}\hline\noalign{\smallskip}
$N_2^{\sigma}$ & 324 &\\
\noalign{\smallskip}\hline\noalign{\smallskip}
$N_2^{\Sigma}$ & 984 &\\
\noalign{\smallskip}\hline
\end{tabular}
\end{table} 

\begin{table}
\caption{
    Degeneracy $g(\Omega,\Sigma)$
   for states of 
   tetrahedron with one \mbox{$S=1/2$} 
   and three \mbox{$S=1$} spins as 
   function of 
   total spin \mbox{$\Omega=\{1/2,3/2,5/2,7/2\}$}, 
   and spin of \mbox{$S=1$} subsystem 
   \mbox{$\Sigma=\{0,1,2,3\}$}, 
    and values of associated coefficients. 
}
\label{table4}
\begin{tabular}{lllll} 
   \hline\noalign{\smallskip}
 & $\Sigma=0$  & $\Sigma =1$ & $\Sigma =2$ & $\Sigma=3$\\
   \noalign{\smallskip}\hline\noalign{\smallskip}
$\Omega=1/2$ & 1 & 1& 1& 1\\
   \noalign{\smallskip}\hline\noalign{\smallskip}
$\Omega=3/2$ & 0& 1& 2& 2\\
   \noalign{\smallskip}\hline\noalign{\smallskip}
$\Omega=5/2$ & 0& 0& 1& 2\\
   \noalign{\smallskip}\hline\noalign{\smallskip}
$\Omega=7/2$ & 0& 0& 0& 1\\
\noalign{\smallskip}\hline\noalign{\smallskip}
\noalign{\smallskip}\hline\noalign{\smallskip}
$N_0$ & 54 &&&\\
\noalign{\smallskip}\hline\noalign{\smallskip}
$N_1$ & 362 &&&\\
\noalign{\smallskip}\hline\noalign{\smallskip}
$N_1^{\Sigma}$ & 468 &&&\\
\noalign{\smallskip}\hline\noalign{\smallskip}
$N_2$ & 3645 &&&\\
\noalign{\smallskip}\hline\noalign{\smallskip}
$N_2^{\Sigma}$ & 3687 &&&\\
\noalign{\smallskip}\hline
\end{tabular}
\end{table} 

\begin{table}
\caption{Degeneracy $g(\Omega)$ for states of 
   spin $S=1$ tetrahedron with total spin $\Omega$ 
  and values of associated coefficients.
}
\label{table5}
\begin{tabular}{llllll} 
\hline\noalign{\smallskip}
      & $\Omega=0$  & $\Omega=1$  & $\Omega= 2$  
      & $\Omega=3$  & $\Omega=4$\\ 
\noalign{\smallskip}\hline\noalign{\smallskip}
S=1   & 3  & 6  &  6  &  3  & 1\\ 
\noalign{\smallskip}\hline\noalign{\smallskip}
\noalign{\smallskip}\hline\noalign{\smallskip}
$N_0$ & 81 & & & &\\
\noalign{\smallskip}\hline\noalign{\smallskip}
$N_1$ & 648 & & & &\\
\noalign{\smallskip}\hline\noalign{\smallskip}
$N_2$ & 776 & & & &\\
\noalign{\smallskip}\hline
\end{tabular}
\end{table} 

\bibliographystyle{unsrt}
\bibliography{paper}

\begin{thebibliography}{10}
\bibitem{note}
We~have checked that includind the constant term $\chi_0$ self consistantly in
  the mean field term does not alter the fits in any~meaningful way.
\end{thebibliography}

\end{document}